\documentclass{elsart}
\usepackage{amsmath,graphicx}
\usepackage{amsmath,euscript,amssymb}
\usepackage[numbers]{natbib}
\newbox\mybox
\newcommand\fverb{\setbox\mybox=\hbox\bgroup\verb}
\newcommand\fverbdo{\egroup\medskip\noindent\fbox{\unhbox\mybox}\ }
\newcommand\fverbit{\egroup\item[\fbox{\unhbox\mybox}]}

\newcommand\init[1]{\setbox\mybox=\hbox{{\beeg #1}~}%
                   \noindent\global\hangindent=\wd\mybox\global\hangafter-2%
                   \sc\smash{\llap {\lower 13.2pt \box\mybox}}}
\def\v1{\vspace{1cm}}
\def\be{\begin{equation}}
\def\ee{\end{equation}}
\def\bc{\begin{center}}
\def\ec{\end{center}}
\def\vh{\varphi}

\newcommand{\bea}{\begin{eqnarray}}
\newcommand{\eea}{\end{eqnarray}}

\begin{document}
\begin{frontmatter}
\title{Conformal Cosmological Model Test with Distant SNIa Data}
\author[china,itep,dub]{A.F. Zakharov,}
\author[msu]{A.A. Zakharova,}
\author[dub]{V.N. Pervushin}
\address[china]{National Astronomical Observatories of Chinese Academy of
Sciences, 20A Datun Road, Chaoyang District, Beijing, 100012, China}
\address[itep]{Institute of Theoretical and Experimental Physics, B. Cheremushkinskaya, 25,
117259, Moscow, Russia}
\address[dub]{Bogoliubov Laboratory for Theoretical Physics, JINR, 141980 Dubna,
Russia}
\address[msu]{Faculty of Mechanics and Mathematics, Moscow State University, Moscow, 119992, Russia}
\begin{abstract}
Assuming that supernovae type Ia (SNe Ia) are standard candles one
could use them to test cosmological theories. The Hubble Space
Telescope team analyzed  186 SNe Ia \cite{Riess_04} to test the
standard cosmological model (SC) and evaluate its parameters. We use
the same sample to determine parameters of Conformal Cosmological
models (CC). We concluded, that really the test is extremely useful
and allows  to evaluate parameters of the model. From a formal
statistical point of view the best fit of the CC model is almost the
same quality approximation as  the best fit of SC model with
$\Omega_\Lambda=0.72, \Omega_m=0.28$. As it was noted earlier, for
CC models, a rigid matter component could substitute the
$\Lambda$-term (or quintessence) existing in the SC model.

\end{abstract}
\end{frontmatter}
%

\section{Introduction}

Now there is enormous progress in observational and theoretical
cosmology and even it is typically accepted that cosmology enters
into an era of precise science (it means that a typical accuracy of
standard parameter determination is about few percents), despite,
there are different approaches including alternative theories of
gravity to fit observational data (see recent reviews
\cite{Will_2006} for references). Some classes of theories could be
constrained by Solar system data \cite{Zakharov_06} even if they
passed cosmological tests. Thus, all the theories should pass all
possible tests including cosmological ones.

Since the end of the last century distant supernovae data is a
widespread test for all theoretical cosmological models in spite of
the fact the correctness of the hypothesis about SNe Ia as the
perfect standard candles is still not proven \cite{Panagia_05}.
However, the first observational conclusion about accelerating
Universe and existence of non-vanishing $\Lambda$-term was done with
the cosmological SNe Ia data.

Therefore, typically  standard (and alternative) cosmological
approaches are checked with the test.

Conformal cosmological models (CC) are also discussed among other
possibilities \cite{CC_papers}. First attempts to analyze SN data to
evaluate parameters of CC models were done \cite{Behnke_02}, so it
was used only 42 high redshift type Ia SNe \cite{Riess_98}, but
after that it was analyzed a slightly extended sample
\cite{Behnke_04}.
 In spite of a small size of the samples used in previous attempts
 to fit CC model parameters, it was
concluded that if $\Omega_{\rm rig}$ is significant in respect to
the critical density, CC models could fit SN Ia observational data
with a reasonable accuracy. An aim of the paper is to check and
clarify previous conclusions about possible bands for CC parameters
with a more extended  (and more accurate) sample \cite{Riess_04}
used commonly to check standard and alternative cosmological models.
The HST cosmological SNe Ia team have corrected data of previous
smaller samples as well and also considered possible
non-cosmological but astronomical ways to fit observational ways and
concluded that some of them such a replenishing dust (with
$\Omega_m=1, \Omega_\Lambda=0.$) could fit observational data pretty
well even in respect to the best fit cosmological model.

The content of the paper
 is the following.
 In Section 2, the basic CC relations are reminded.
 In Section 3, a magnitude-redshift relation for distant SNe is
 discussed. In Section 4,
 results of fitting procedure for CC models with the "gold" and "silver" sample and the  "gold" subsample
 are given.  Conclusions are presented in Section 5.

\section{\label{s-1} Conformal Cosmology Relations}
We will remind basic relations between observational data  and CC
model parameters.  The correspondence between the SC and the CC is
determined by the evolution of the dilaton \cite{Behnke_02} \be
\label{cce} (\vh')^2 = \rho(\vh)~, \ee where the prime denotes the
derivative with respect to the conformal time $\eta$ and the time,
the density $ \rho(\vh)$, and the Hubble parameter $H_0$ are treated
as measurable quantities. The standard cosmological definitions of
the redshift and the density parameter are the following \bea
\label{denspar}
1+z \equiv \frac{1}{a(\eta)}=\frac{\vh_0}{\vh(\eta)}~,
~~~\Omega(z)=\frac{\rho(\vh)}{\rho(\vh_0)} , \eea where
$\Omega(0)=1$ is assumed. The density parameter $\Omega(z)$ is
determined   \cite{Behnke_02} \be \label{coc} \Omega(z)=\Omega_{\rm
rig}(1+z)^2+\Omega_{\rm rad}+ \frac{\Omega_{m}}{(1+z)} +
\frac{\Omega_{\Lambda}}{(1+z)^4}~. \ee We note here that all the
equations of state that are known in the standard cosmology: the
rigid state ($p_{\rm rig} = \rho_{\rm rig}(\vh)= {\rm
const}/\vh^2$), the radiation state ($p_{\rm rad} = \rho_{\rm rad} /
3 = {\rm const}$), and the matter state ($p_{m} = 0, ~~\rho_{m} =
{\rm const} \cdot \vh $)~\cite{ps1,pp}.

 Then the equation~(\ref{cce}) takes the form
\be \label{etad}
H_0\frac{d\eta}{dz}=\frac{1}{(1+z)^2}\frac{1}{\sqrt{\Omega(z)}}~,
\ee and determines the dependence of the conformal time on the
redshift factor. This equation is valid also for the conformal time
- redshift relation in the SC where this conformal time is used for
description of a light ray.

A light ray traces a null geodesic, i.e. a path for which the
conformal interval $(ds^L)^2=0$ thus satisfying the equation
${dr}/{d\eta} = 1$. As a result we obtain for the coordinate
distance as a function of the redshift \be \label{rdi} H_0
r(z)=\int_0^z \frac{dz'}{(1+z')^2}\frac{1}{\sqrt{\Omega(z')}}. \ee
The equation~(\ref{rdi}) coincides with the similar relation between
coordinate distance and redshift in SC.

In the comparison with the stationary space in SC and stationary
masses in CC, a part of photons is lost. To restore the full
luminosity in both SC and CC we should multiply the coordinate
distance by the factor $(1+z)^2$. This factor comes from the
evolution of the angular size of the light cone of emitted photons
in SC, and from the increase of the angular size of the light cone
of absorbed photons in CC.

However, in  SC  we have an additional factor $(1+z)^{-1}$ due to
the expansion of the universe, as measurable distances in SC are
related to measurable distances in CC (that coincide with the
coordinate ones) by the relation \be \ell={a}\int\limits_{ }^{
}\frac{dt}{a}=\frac{r}{1+z}~.
\ee Thus, we obtain the relations \be \label{scr} \ell_{\rm SC}(z) =
(1+z)^2 \ell = (1+z)  r(z)~, \ee \be \label{ccr} \ell_{\rm CC}(z) =
(1+z)^2  r(z)~. \ee This means that the observational data are
described by different regimes in SC and CC.

\section{\label{s-2} Magnitude-Redshift Relation}

Typically to test cosmological theories one should check a relation
between an apparent magnitude and a redshift. In both SC and CC
models it should be valid the effective magnitude-redshift relation:
\be \mu (z)\equiv m(z)- M = 5 \log{[H_0\ell(z)]} + {\mathcal M},
\label{magz} \ee
 where
$m(z)$ is an observed magnitude, $M$ is the absolute magnitude,
${\mathcal M}$ is a constant with recent experimental data for
distant SNe. Values of $\mu_i$, $z_i$ and $\sigma_i$ could be taken
from observations of a detected supernova with index $i$
($\sigma_i^2$ is a dispersion for the $\mu_i$ evaluation). Since we
deal with observational data  we should choose model parameters to
satisfy an array of relations (\ref{magz}) by the best way because
usually, a number of relations is much more than a number of model
parameters and there are errors in both theory and observations (as
usual we introduce indices for the relations corresponding to all
objects). Typically, $\chi^2$-criterium is used to solve the
problem, namely, we calculate \be \chi^2 = \sum_i \frac{(\mu_i^{\rm
theor}-\mu_i)^2}{\sigma_i^2}, \label{chi2} \ee where $\mu_i^{\rm
theor}$ are calculated for given $z_i$ with the assumed theoretical
model and after that we can evaluate the best fit model parameters
minimizing $\chi^2$-function.

\section{\label{s-2} Model Fits}

\subsection{Total sample analysis}

For the standard cosmological model for the 186 SNe (the "gold" and
"silver" sample),\footnote{To express differences in quality of
spectroscopic and photometric data the supernovae were separated
into "high-confidence" ("gold") and "likely but not certain"
("silver") subsets \cite{Riess_04}.}  a minimum of the
$\chi^2$-function gives us $\Omega_m=0.28$ ($\chi_{\rm
SC~flat}^2=232.4$) and $\Omega_m=0.31, \Omega_\Lambda=0.80$ assuming
$|\Omega_k| \leqslant 0.11$ ($\chi_{\rm SC~flat}^2=231.0$). Since
other cosmological tests dictate that the Universe should be almost
flat and $\Omega_m=0.28$ is an acceptable value \cite{Will_2006}, we
choose the flat SC model for a reference.

\begin{figure}[t!]
\begin{center}
\includegraphics[width=10.5cm]{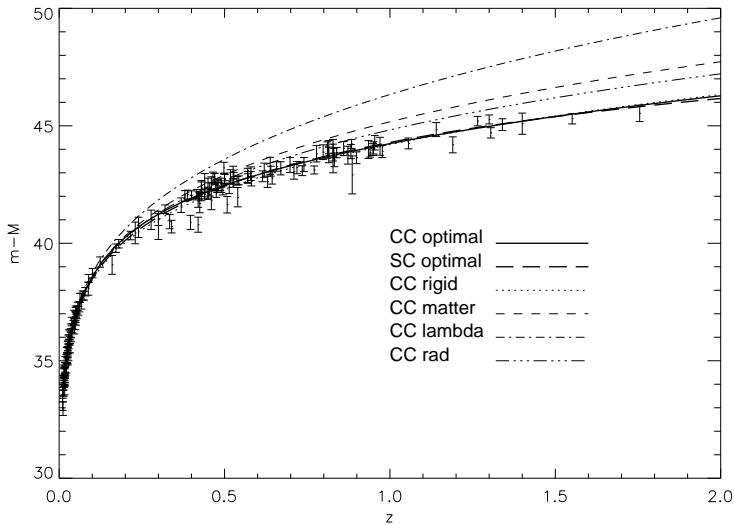}
\end{center}
\caption{$\mu(z)$-dependence for  cosmological models in SC and CC.
The data points include 186 SN Ia (the "gold" and "silver" sample)
used by the cosmological supernova HST team \cite{Riess_04}. For a
reference we use the best fit for the flat standard cosmology model
with $\Omega_m=0.27, \Omega_\Lambda=0.73$ (the thick dashed line),
the best fit for CC is shown with the thick  solid line. For this CC
model we do not put any constraints on $\Omega_m$.} \label{fig1}
\end{figure}

\begin{table}[t!]
\begin{center}
\begin{tabular}{|c|c|c|c|c|c|}
\hline
 Constraints on $\Omega_m$ & $\Omega_m$ & $\Omega_\Lambda$ & $\Omega_{\rm rad}$ & $\Omega_{\rm rig}$ & $\chi^2$\\
\hline \hline
No constraints             & -4.13      & 3.05             & 0.05               & 2.085              & 226.64 \\
\hline
$\Omega_m \ge 0.$          & 0.         & 0.18             & 0.                 & 0.80               & 242.76 \\
\hline
$0.2 \le \Omega_m \le 0.3$ & 0.2        & 0.013            & 0.                 & 0.75               & 244.67\\
 \hline
$0.2 \le \Omega_m \le 0.3$ & 0.29        & 0.0            & 0.                 & 0.7               & 246.58\\
 \hline
 $0.2 \le \Omega_m \le 0.3$ & 0.27        & 0.0            & 0.                 & 0.72               & 245.66\\
 \hline
\end{tabular}
\caption{The fits for CC models for the total sample with different
constraints on $\Omega_m$ (the best fits are shown in first, second
and third rows, two almost best fits are presented in fourth and
fifth rows).}
 \end{center}
\label{tabl1}
\end{table}

\begin{table}[t!]
\begin{center}
\begin{tabular}{|c|c|c|c|c|c|}
\hline
Model types  &$\Omega_m=1$ & $\Omega_\Lambda=1$ & $\Omega_{\rm rad}=1$ & $\Omega_{\rm rig}=1$ \\
\hline \hline
$\chi^2$        & 924.27      & 4087.93             & 478.42               & 276.71      \\
\hline
 \end{tabular}
\caption{The $\chi^2$ values for pure flat CC models for the total
sample. The models are shown in Figs. \ref{fig1},\ref{fig2} as
references.}
 \end{center}
\label{tabl2}
\end{table}

In Fig.~\ref{fig1} we compare the SC and CC fits  for the effective
magnitude-redshift relation if we will not put any constraint on
$\Omega_m$ (in this case we assume that SNe Ia data is the only
cosmological test for CC models we obtain the best fit expressed in
the first row in Table~\ref{tabl1}). Analyzing the curves
corresponding to the best fits for SC and CC models one can say that
they almost non-distinguishable, moreover the best fit CC provide
even better the $\chi^2$ value (see first row in Table \ref{tabl1}).
We would not claim that we discovered a cosmological model with
negative $\Omega_m$, but we would like to note that the best CC and
SC fits are almost non-distinguishable from a formal statistical
point of view (the thick solid and long dashed lines, respectively
in Fig.~\ref{fig1}). Sometimes new physical phenomena are introduced
qualitatively with  the same statistical arguments (such as an
introduction of the phantom energy, for instance), but if we should
follow a more conservative approach, we could conclude that in this
case we should simply put extra constraints on $\Omega_m$ to have no
contradictions to other cosmological (and astronomical) tests.
 So,
if we put "natural" constraints on $\Omega_m \geqslant 0$, the best
fit parameters for CC model are presented in second row in Table
\ref{tabl1}. In this case the $\chi^2$ difference between two CC
models ($\Delta \chi^2 \thickapprox 16$) is not very high and a
difference between this fit and the SC best fit for a flat model is
about $\Delta \chi^2 \thickapprox 10$ (or less than 5\%), it means
the CC fit is at an acceptable level. For references, we plotted
also pure flat CC models, so that rigid, matter, lambda and
radiation models are shown with thin dotted, short dashed, dot dash,
dash dot dot dot lines, respectively. Corresponding $\chi^2$ values
are given in Table~2. One can see that only pure flat rigid CC model
has relatively low $\chi^2$ values (and it could be accepted as a
rough and relatively good fit for cosmological SNe Ia data), but
other models should be definitely ruled out by the observational
data.

\begin{figure}[t!]
\begin{center}
\includegraphics[width=10.5cm]{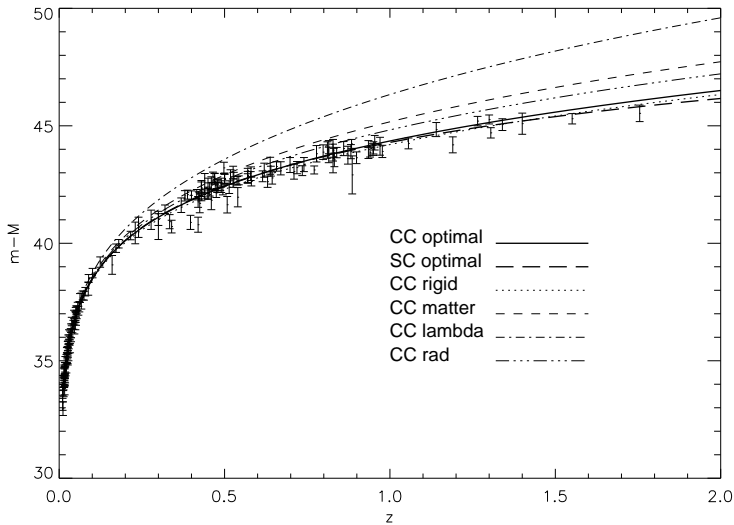}
\end{center}
\caption{$\mu(z)$-dependence  cosmological models in SC and CC. As
in previous figure, the data points include 186 SN Ia (the "gold"
and "silver" sample) used by the cosmological supernova HST team
\cite{Riess_04} and for a reference we use the best fit for the flat
standard cosmology model with $\Omega_m=0.27, \Omega_\Lambda=0.73$
(the thick dashed line), the best fit for CC is shown with the thick
solid line. For this CC model we assume $\Omega_m \in [0.2,0.3]$.}
\label{fig2}
\end{figure}

So, if we put further constraints on $0.2 \leqslant \Omega_m
\leqslant 0.3$ based on measurements of clusters of galaxies and
other cosmological arguments \cite{Bahcall_98}, the best fit
parameters for CC model are presented in third row in Table~1. In
this case the $\chi^2$ difference between two CC models ($\Delta
\chi^2 \thickapprox 18$) is not very high also and a difference
between $\chi^2$ for the CC  and  SC models is about $\Delta \chi^2
\thickapprox 12$ (or about 5\%), it means the CC fit is at an
acceptable level. Dependence of $\chi^2$ on $\Omega_m$ is very weak
and we present intermediate fits for CC model in fourth and fifth
rows in Table~1 (there is a valley of $\chi^2$ function in the
$\Omega_m$ direction). The best fit for a CC model with parameters
given in third row in Table~1 is shown as the optimal fit for the CC
model in Fig.~\ref{fig2} with the solid thick line. Other lines are
the same as in Fig.~\ref{fig1} and they are shown for references.

\subsection{Analysis of the "Gold" Subset}

We also did the same calculations for "gold" subset of SNe Ia data
(157 objects). The best fit for SC  flat model corresponds to
$\Omega_m=0.285$ and $\chi^2=177.10$ (if we try to find the best fit
for SC model with the constraint $|\Omega_k| \leqslant 0.11$ we have
$\Omega_m=0.32$, $\Omega_\Lambda=0.79$ and $\chi^2=176.07$), however
as in the previous case for the total sample, we have selected
generally accepted the SC flat model for a reference with
$\Omega_m=0.285$.

\begin{figure}[t!]
\begin{center}
\includegraphics[width=10.5cm]{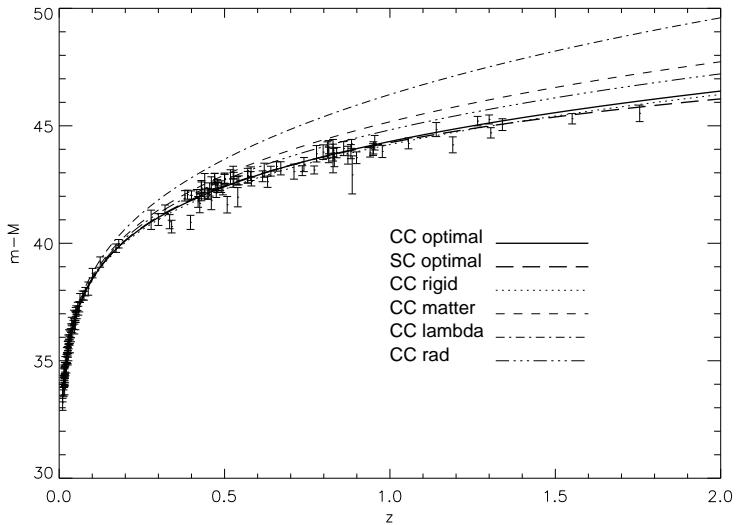}
\end{center}
\caption{$\mu(z)$-dependence for  cosmological models in SC and CC.
The data points include 157 SN Ia (the "gold" subsample) used by the
cosmological supernova HST team \cite{Riess_04}. For a reference we
use the best fit for the flat standard cosmology model with
$\Omega_m=0.285, \Omega_\Lambda=0.715$ (the thick dashed line), the
best fit for CC is shown with the thick  solid line.  For this CC
model we do not put any constraints on $\Omega_m$.} \label{fig3}
\end{figure}

The parameters for the best fits for CC models are given in first
row in Table~3 (the corresponding plot is shown in Fig. \ref{fig3}).
One can note here that in spite of the previous case for the "gold"
and "silver" sample the best fit for CC model does not need so
exotic negative $\Omega_m$. Analyzing $\chi^2$ presented in Table~3
(last column), one can say that like in the previous case,
dependence of $\chi^2$ function on $\Omega_m$ is very weak and SNe
Ia data is not very good way to evaluate precisely $\Omega_m$ for CC
models.

\begin{table}[t!]
\begin{center}
\begin{tabular}{|c|c|c|c|c|c|}
\hline
 Constraints on $\Omega_m$ & $\Omega_m$ & $\Omega_\Lambda$ & $\Omega_{\rm rad}$ & $\Omega_{\rm rig}$ & $\chi^2$\\
\hline \hline
No constraints             & .16      &  0.0             & 0.0               & 0.76              & 187.20 \\
\hline
$0.2 \le \Omega_m \le 0.3$ & 0.2        & 0.0            & 0.                 & 0.74               & 187.50\\
 \hline
 $0.2 \le \Omega_m \le 0.3$ & 0.264        & 0.0            & 0.                 & 0.73               & 188.94\\
 \hline
\end{tabular}
\caption{The fits for CC models for the "gold" subsample with
different constraints on $\Omega_m$ (the best fits are shown in
first, second rows, a good fit is presented in third row).}
 \end{center}
\label{tabl3}
\end{table}

\begin{table}[t!]
\begin{center}
\begin{tabular}{|c|c|c|c|c|c|}
\hline
Model types  &$\Omega_m=1$ & $\Omega_\Lambda=1$ & $\Omega_{\rm rad}=1$ & $\Omega_{\rm rig}=1$ \\
\hline \hline
$\chi^2$        & 809.79      & 3631.55             & 407.02               & 210.0      \\
\hline
 \end{tabular}
\caption{The $\chi^2$ values for pure flat CC models for the the
"gold" subsample. The models are shown in
Figs.~\ref{fig3},\ref{fig4} as references.}
 \end{center}
\label{tabl4}
\end{table}

\begin{figure}[t!]
\begin{center}
\includegraphics[width=10.5cm]{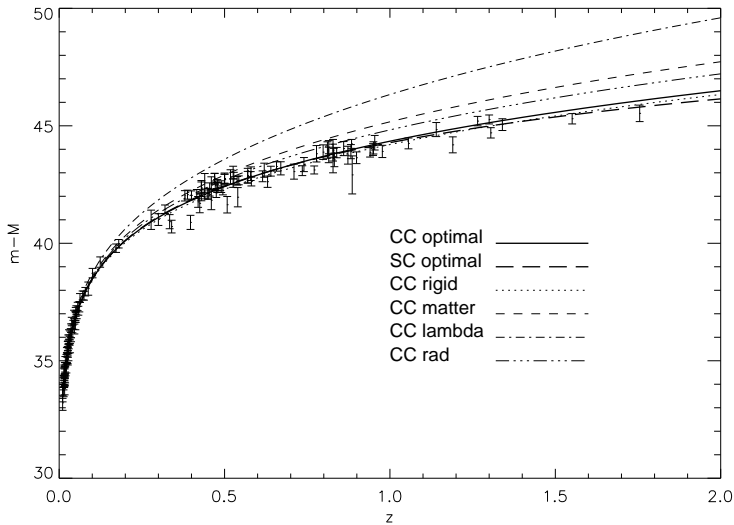}
\end{center}
\caption{$\mu(z)$-dependence for  cosmological models in SC and CC.
The data points include 157 SN Ia (the "gold" subsample). For a
reference we use the best fit for the flat standard cosmology model
with $\Omega_m=0.285, \Omega_\Lambda=0.715$ (the thick dashed line),
the best fit for CC is shown with the thick solid line. For this CC
model we assume $\Omega_m \in [0.2,0.3]$.} \label{fig4}
\end{figure}

As for the total sample, we obtain that for the "gold" subset a
$\chi^2$ difference between $\chi^2$ for the CC and SC models is
about $\Delta \chi^2 \thickapprox 12$ (or about 5\%), it means the
CC fit is at an acceptable level, however, the best SC flat model is
still a little bit more preferable.

\section{\label{s-3}Conclusions}

Using "gold" and "silver" 186 SNe Ia \cite{Riess_04} we confirm in
general and clarify previous conclusions about CC model parameters,
done earlier with analysis of smaller sample of SNe Ia data
\cite{Behnke_02, Behnke_04} that the pure flat rigid CC model could
fit the data relatively well since $\Delta \chi^2 \thickapprox 44.3$
(or less than 20~\%) in respect of the standard cosmology flat model
with $\Omega_m=0.28$. Other pure flat CC models should be ruled out
since their $\chi^2$ values are too high.

For the total sample, if we consider CC models with a "realistic"
constraint $0.2 \leqslant \Omega_m \leqslant 0.3$ based on other
astronomical or cosmological arguments except SNe Ia data, we
conclude that the standard cosmology flat model with $\Omega_m=0.28$
is still preferable in respect to the  fits for the CC models (with
$\Omega_m=0.2$ and $\Omega_{\rm rig}=0.75$ or $\Omega_m=0.27$ and
$\Omega_{\rm rig}=0.72$, for instance, see third and fifth rows in
Table~1),  but the preference is not very high (about 5~\% in
relative units of $\chi^2$ value), so the CC models could be adopted
as acceptable ones taking into account possible sources of errors in
the sample and systematics.

With the presented analysis for the "gold" subsample  we re-confirm
results given with the total sample except the absence of the best
fit for CC models with the negative $\Omega_m$ as it was obtained
with the total sample. The best fits for CC models without
constraints and with "natural" constraints on $\Omega_m$ such as
$\Omega_m \in [0.2,0.3]$ give almost the same  curves for
magnitude-redshift dependences. For the "gold" subsample, a $\chi^2$
difference is about then 5~\%  for the best fit CC and SC models, it
means that the conclusion obtained earlier for the total ("gold" and
"silver") sample is also correct for the "gold" subsample.

Thus, for  CC model fits calculated with SNe Ia data, in some sense,
a rigid equation of state could substitute the $\Lambda$-term (or
quintessence) in the Universe content.

The best CC models provide almost the same quality fits of SNe Ia
data as the best fit  for the SC flat model, however the last
(generally accepted) model is more preferable.


\section*{Acknowledgements}

The authors are indebted to B.M.~Barbashov, J.~Wang, J.~Zhang,
G.~Zhou for important discussions.
 AFZ is grateful to the National Natural
Science Foundation of China (NNSFC)  (Grant \# 10233050) and
National Basic Research Program of China (2006CB806300) for a
partial financial support of the work.

\end{document}